\documentclass[12pt]{article}
\usepackage{epsfig,amsmath,amssymb,mathrsfs}
\usepackage[utf8]{inputenc}
\usepackage[colorlinks=true,citecolor=blue,,linktocpage=true,linkcolor=blue,urlcolor=black]{hyperref}
\usepackage{mathtools,slashed}
\usepackage{array}

\usepackage[force]{feynmp-auto}
\usepackage{caption}
\usepackage{subcaption}

  {\end{list}}%

\makeatletter
\renewcommand{\section}{\setcounter{equation}{0}\@startsection
 {section}%
 {1}%
 {0pt}%
 {-1\baselineskip}%
 {0.4\baselineskip}%
 {\bfseries\large}}%
\renewcommand{\subsection}{\@startsection
 {subsection}%
 {2}%
 {0pt}%
 {-0.75\baselineskip}%
 {0.2\baselineskip}%
 {\bfseries}}%
\renewcommand{\subsubsection}{\@startsection
 {subsubsection}%
 {3}%
 {0pt}%
 {-0.5\baselineskip}%
 {0.1\baselineskip}%
 {\sc}}%
\makeatother

\DeclareMathAlphabet{\mathpzc}{OT1}{pzc}{m}{it}

\def\be{\begin{equation}}
\def\ee{\end{equation}}
\def\a{\alpha}

\def\g5{\gamma_{5}}

\def\m{\mu}
\def\n{\nu}

\def\Aslash{{A\mkern-11mu/}}

\def\kslash{{k\mkern-8mu/}{\!}}
\def\prslash{{\partial\mkern-9mu/}}

\def\prslash{{\partial\mkern-9mu/}}    %%_standard_Dirac_operator

\def\id3k{\int\!\! \dfrac{d^3\!\vec{k}}{(2\pi)^3 2E(\vec{k})}}

\def\idx{\int\!\! d^4\!x}

\def\idy{\int\!\! d^4\!y}

\def\idx{\int\!\! d^{4}\!x}
\newcommand{\bea}{\begin{eqnarray}}
\newcommand{\eea}{\end{eqnarray}}
\newcommand{\beann}{\begin{eqnarray*}}
\newcommand{\eeann}{\end{eqnarray*}}
\newcommand{\ba}{\begin{array}}
\newcommand{\ea}{\end{array}}

 \def\g {\gamma}
\newcommand{\email}[1]{\href{mailto:#1}{\tt #1}}

\begin{document}

\rightline{\scriptsize{FTI/UCM 217-2019}}
\vglue 50pt

\begin{center}

{\LARGE \bf Off-shell unimodular $N=1, d=4$ supergravity}\\
\vskip 1.0true cm
{\Large Jesus Anero$^{\dagger}$, Carmelo P. Martin$^{\ddag}$ } {\large and} {\Large Raquel Santos-Garcia$^{\dagger}$}
\\
\vskip .7cm
{
	$^{\dagger}$Departamento de F\'isica Te\'orica and Instituto de F\'{\i}sica Te\'orica (IFT-UAM/CSIC),\\
	Universidad Aut\'onoma de Madrid, Cantoblanco, 28049, Madrid, Spain\\
	\vskip .1cm
	$^{\ddag}${Universidad Complutense de Madrid (UCM), Departamento de Física Teórica and IPARCOS, Facultad de Ciencias Físicas, 28040 Madrid, Spain}
	
	\vskip .5cm
	\begin{minipage}[l]{.9\textwidth}
		\begin{center}
			\textit{E-mail:}
			\email{jesusanero@gmail.es},
			\email{carmelop@fis.ucm.es}, \email{raquel.santosg@uam.es}
			
		\end{center}
	\end{minipage}
}
\end{center}
\thispagestyle{empty}

\begin{abstract}
We formulate a unimodular $N=1, d=4$ supergravity theory off shell. We see that the infinitesimal Grassmann parameters defining the unimodular supergravity transformations are constrained
and show that the conmutator of two infinitesinal unimodular supergravity transformations closes on transverse diffeomorphisms, Lorentz transformations and unimodular supergravity transformations.
Along the way, we also show that the linearized theory is a supersymmetric theory of gravitons and gravitinos. We see that de Sitter and anti-de Sitter spacetimes are non-supersymmetric vacua of our
unimodular supergravity theory.

\end{abstract}

%\vspace{9pt}
\vfill
\clearpage

\section{Introduction}

As a classical theory, unimodular gravity is a geometric theory of gravity obtained by defining the configuration space of theory to be the set of Lorentzian metrics which satisfy the unimodular condition $\det\,g_{\mu\nu}=-1$. Hence, the covariance group of theory is no longer the group of diffeomorphisms but a subgroup of it: the group of transverse diffeomorphisms. As far as the classical equations of motion are concerned, unimodular gravity cannot be distinguised from Einstein's general relativity with an arbitrary Cosmological Constant. Unimodular gravity thus appears to be a viable classical theory of gravity --for more information, see \cite{Ellis:2010uc, Ellis:2013eqs}.
That the determinant of the metric is no longer a degree of freedom has the consequence that the vacuum energy does not gravitate. And thus, the
problem that arises in General Relativity of the huge discrepancy between the experimental value of the Cosmological Constant and the quantum field theory "prediction"  for that constant does occur in unimodular gravity \cite{Weinberg:1988cp}. Of course, unimodular gravity does not predict the value of the Cosmological Constant.

Although unimodular gravity and General Relativity seem to be equivalent classically, this is not so at the quantum level, at least when analyzing phenomena where the  Cosmological Constant cannot be set to zero. Some properties of unimodular gravity defined as an effective quantum field theory have been analysed in a number of papers. It all points in the direction that when the Cosmological Constant is set to zero, unimodular gravity and General relativity have the same S matrix in the perturbative regime, but a proof of this statement is still lacking. We refer the reader to  References \cite{Smolin:2009ti}--\cite{deBrito:2019umw} for further information.

Supergravity was introduced in Reference \cite{Freedman:1976xh} and it is difficult to overstate the impact that it has had and still has on high energy physics --see Reference \cite{Freedman:2012zz} for a modern  introduction. One of the salient, and most surprising, features of supergravity is the  way it ushers in Grassmann variables as a key ingredient in the description of the spacetime dynamics. It thus seems imperative to see whether unimodular gravity can be supersymmetrized to a supergravity theory.

The purpose of this paper is to formulate the off-shell $N=1, d=4$ Poincar\'e supergravity counterpart of unimodular gravity. We shall call this theory $N=1, d=4$ unimodular supergravity.
This is not the first time in the literature that a supergravity counterpart of unimodular gravity is proposed. A unimodular supergravity was put forward in reference \cite{Nishino:2001gd}. Our unimodular  supergravity formalism differs from the one in reference \cite{Nishino:2001gd} by three main aspects. First, our formalism is off-shell, theirs is on-shell. It should be noticed that the existence of an off-shell formulation of a supersymmetric theory is a highly non-trivial issue -see Reference \cite{Gates:2001rn}. Second, we do not use any Langrange multiplier to implement the vierbein unimodularity condition, so we deal with a minimum number of fields. Third, the equation to be satisfied by the parameters of the unimodular supergravity transformations is not the same as in reference  \cite{Nishino:2001gd}.

The layout of this paper is the following. In section 2 we put forward the linearized $N=1, d=4$ unimodular supergravity theory with auxiliary fields and show that the fields carry an off-shell representation of $N=1$ supersymmetry in four dimensions, up to gauge transformations. We also see that it is a theory of free gravitons and gravitinos. The off-shell interacting unimodular supergravity theory of gravitons and gravitinos is put forward in section 3. Here, we show that the algebra of unimodular supergravity transformations closes modulo transverse diffeomorphisms and Lorentz transformations. This closing is non trivial since on the one hand the parameters defining the unimodular supergravity transformations are constrained and only transverse diffeomorphisms are allowed as symmetries. In section 4, we  discuss several aspects of the classical solutions to the unimodular gravity equations of motion.

\section{Linearized $N=1, d=4$ unimodular supergravity}

In this section we shall supersymmetrize the linearized unimodular gravity theory. The action, $S^{(LUG)}$, of the latter is obtained \cite{Alvarez:2005iy, Alvarez:2006uu} by imposing the tracelessness constraint --the linearized unimodular condition,
\begin{equation}
h_{\mu}^{\mu}=0,
\label{tracelessness}
\end{equation}
on the graviton field in the  Fierz-Pauli action. Thus, one obtains
\begin{equation}
S^{(LUG)}\,=\,-\frac{1}{4}\idx\,\Big[\frac{1}{2}\,h^{\mu\nu}\partial^2 h^{\mu\nu}\,+\,\partial_{\mu}h^{\mu\lambda}\partial_{\nu}h^{\nu}_{\phantom{\nu}\lambda}\Big].
\label{SLUG}
\end{equation}
Note that $\partial^2$ stands for $\partial_{\mu}\partial^{\mu}$.

$S^{(LUG)}$ is not invariant under arbitrary infinitesimal diffeormophisms but only under transverse infinitesimal diffeomorphisms:
\begin{equation}
\delta_{Tdiff}h_{\mu\nu}\,=\,\partial_{\mu}\xi_{\nu}\,+\,\partial_{\nu}\xi_{\mu}\quad\text{with}\quad \partial_{\mu}\xi^{\mu}=0.
\label{Tdiff}
\end{equation}

It is plain that to have a chance of obtaining a supersymmetric theory whose particle spectrum contains the graviton field, $h_{\mu\nu}$, we must add to the action $S^{(LUG)}$ the Rarita-Schwinger action, $S^{(RS)}$, for a spin-$3/2$ Majorana spinor field $\psi_{\mu}$:
\begin{equation}
S^{(RS)}\,=-\,\frac{i}{2}\idx\,\overline{\psi}_{\mu}\gamma^{\mu\nu\rho}\partial_{\nu}\psi_{\rho},
\label{RSaction}
\end{equation}
where $\gamma^{\mu\nu\rho}=\gamma^{[\mu}\gamma^{\nu}\gamma^{\rho]}$. $[\mu\nu\rho]$ stands for total antisymmetrization of the indices with weight 1, ie, there is a global factor $1/3!$ multiplying the sum running over all the signed permutations of $(\mu,\nu,\rho)$. Note that $\epsilon^{0123}=1$.

The action, $S^{(RS)}$, is invariant under the following gauge transformations
\begin{equation}
\delta_{gauge}\psi_{\mu}=\partial_\mu\chi,
\label{gtrans}
\end{equation}
where $\chi$ is an arbitrary Majorana spinor. Let us recall that each component of the Majorana vector-spinor $\psi_{\mu}(x)$ must be an odd element of a Grassmann algebra to prevent $S^{(RS)}$ from vanishing.

We shall now look for infinitesimal rigid --ie, not dependent on the coordinates-- fermionic transformations which turn the $\psi_{\mu}$ field into the $h_{\mu\nu}$ field, and viceversa, while leaving invariant the  sum $S^{(LUG)}+S^{(RS)}$. An educated guess reads
\begin{equation}
\begin{array}{l}
{\delta_{\epsilon} h_{\mu\nu}\,=\,-\frac{i}{2}\,\overline{\epsilon}(\gamma_{\mu}\psi_{\nu}\,+\,\gamma_{\nu}\psi_{\mu})}\\[4pt]
{\delta_{\epsilon} \psi_{\mu}\,=\,-\frac{1}{4}\,\partial^{\rho}h^{\sigma}_\mu\,\gamma_{\rho\sigma}\epsilon},
\end{array}
\label{firstattempt}
\end{equation}
where $\epsilon$ is an infinitesimal rigid Majorana spinor.

Unfortunately, the previous transformations do not preserve the  unimodular gravity condition in (\ref{tracelessness}), for, in general, $\gamma^{\mu}\psi_{\mu}$ does not vanish. Notice that one cannot take advantage of the invariance of $S^{(LUG)}$ under the transverse diffeomorphisms in (\ref{Tdiff}) and add to $\delta_{\epsilon} h_{\mu\nu}$ above a transverse diffeomorphism so that the unimodular gravity condition, $h^{\mu}_{\mu}=0$ be preserved, for that would entail the following constraint on $\xi_\mu$:
\begin{equation*}
\partial_\mu \xi^{\mu}\,=\,-\overline{\epsilon}\gamma^{\mu}\psi_{\mu},
\end{equation*}
which clashes with the transversality constraint on $\xi^{\mu}$; unless, of course, $\gamma^{\mu}\psi_{\mu}=0.$

It would appear, in view of the previous analysis, that the value of the Majorana vector-spinor field is to be restricted by
constraint
\begin{equation}
\gamma^{\mu}\psi_{\mu}=0,
\label{RSgauge}
\end{equation}
if the $N=1$ supersymmetrization of our linear unimodular gravity theory is to be successful. From now on  we shall always assume that the Majorana spin-$3/2$ field $\psi_{\mu}$ satisfies (\ref{RSgauge}). Notice that (\ref{RSgauge}) is the so-called Rarita-Schwinger gauge, which was introduced in the seminal paper by Rarita and Schwinger \cite{Rarita:1941mf} to characterize the pure spin-3/2 states. In the context of supersymmetry, (\ref{RSgauge}) could be viewed as the supersymmetry counterpart of the tracelessness constraint -see (\ref{tracelessness})--  on the graviton field which defines linear unimodular gravity.

However, we are not done yet, for the constraint $\gamma^{\mu}\psi_{\mu}=0$ is not preserved by the second transformation in (\ref{firstattempt}). Fortunately, now we can add to the right hand side of (\ref{firstattempt}) a suitably chosen gauge transformation, as defined in (\ref{gtrans}), so that the new transformation preserves the constraint $\gamma^{\mu}\psi_{\mu}=0$. We shall not give yet the value of such  gauge transformation, for it is time that we introduce the bosonic auxiliary fields that will lead to the off-shell realization of N=1 supersymmetry.

Let $S$ and $P$ be scalar and pseudoscalar real fields, respectively, and $A_{\mu}$ a real pseudovector field. Guided by supersymmetry transformations of the standard linearized $N=1, d=4$ supergravity theory -see \cite{Ferrara:1978em}, we define the following transformations of the fields of the linearized unimodular supergravity theory under construction
\begin{equation}
\begin{array}{l}
{\delta^{L}_{\epsilon} h_{\mu\nu}\,=\,-\frac{i}{2}\,\overline{\epsilon}(\gamma_{\mu}\psi_{\nu}\,+\,\gamma_{\nu}\psi_{\mu}),}\\[4pt]
{\delta^{L}_{\epsilon} \psi_{\mu}\,=\,-\frac{1}{4}\,\partial^{\rho}h^{\sigma}_\mu\,\gamma_{\rho\sigma}\epsilon+\frac{i}{6}\gamma_{\mu}(S-i\gamma_5 P)\epsilon+\frac{i}{2}\gamma_5 (A_{\mu}-\frac{1}{3}\gamma_{\mu}\gamma^{\nu}A_{\nu})\epsilon+\partial_{\mu}\theta[\epsilon],}\\[4pt]
{\delta^{L}_{\epsilon} S\,=\,-\frac{1}{2}\,\overline{\epsilon}\gamma^{\rho\sigma}\partial_{\rho}\psi_{\sigma},}\\[4pt]
{\delta^{L}_{\epsilon} P\,=\,\frac{i}{2}\,\overline{\epsilon}\gamma_{5}\gamma^{\rho\sigma}\partial_{\rho}\psi_{\sigma},}\\[4pt]
{\delta^{L}_{\epsilon} A_{\mu}\,=\,\frac{3}{4}\overline{\epsilon}\gamma_{5}(\eta_{\mu\rho}-\frac{1}{3}\gamma_{\mu}\gamma_{\rho})
\gamma^{\rho\sigma\lambda}\partial_{\sigma}\psi_{\lambda},}
\end{array}
\label{susytrans}
\end{equation}
where
 $\gamma_{5}=i\gamma^0\gamma^1\gamma^2\gamma^3$
 and
\begin{equation}
\begin{array}{l}
{\theta[\epsilon](x)=-\idy\,D(x-y)\big[-\frac{1}{4}\,\gamma^\mu\partial^{\rho}h^{\sigma}_\mu(y)\,\gamma_{\rho\sigma}\epsilon+\frac{2i}{3}(S(y)-i\gamma_5 P(y))\epsilon+\frac{i}{6}\gamma_5 \gamma^{\mu}A_{\mu}(y)\epsilon\big],}\\[4pt]
{\prslash_{x}D(x-y)=\delta(x-y)}\\[4pt]
{\prslash \theta[\epsilon]=\frac{1}{4}\,\gamma^\mu\partial^{\rho}h^{\sigma}_\mu\,\gamma_{\rho\sigma}\epsilon-\frac{2i}{3}(S-i\gamma_5 P)\epsilon-\frac{i}{6}\gamma_5 \gamma^{\mu}A_{\mu}\epsilon}
\end{array}
\label{compensationgaugetrans}
\end{equation}
The summand $\partial_{\mu} \theta(x)$ in (\ref{susytrans}) is needed so that
\begin{equation}
\delta^{L}_{\epsilon}(\gamma^{\mu} \psi_{\mu})=0,\quad \forall h_{\mu\nu},\psi_\mu,S,P, A_\mu.
\label{vanishingvarlin}
\end{equation}
Some comments regarding $\partial_{\mu} \theta(x)$ are in order. First, it is plain that $\partial_{\mu} \theta(x)$ does not contribute to the variation of the Rarita-Schwinger action in (\ref{RSaction}). Secondly, when it acts on  a physical gravitino it has to be contracted with the corresponding vector-spinor wave function, which is transverse, thus yielding a vanishing constribution.

Let us stress now that (\ref{vanishingvarlin}) does not further constrain the fields $h_{\mu\nu}$ and $\psi_\mu$ since the last equation of  (\ref{compensationgaugetrans}) holds for arbitrary  $h_{\mu\nu},\psi_\mu,S,P,A_\mu$ with appropriate regularity and boundary behaviour.

Recall that in the unimodular theory $h_{\mu\nu}$ and $\psi_{\mu}$ are constrained by the unimodular conditions in (\ref{tracelessness}) and (\ref{RSgauge}), respectively; which are indeed preserved by the transformations in ({\ref{susytrans}). We shall see in the next section that $\delta^{L}_{\epsilon} \psi_{\mu}$ and all the remaining supersymmetry transformations in (\ref{susytrans}) are the order $\kappa^0$  contributions to the  supergravity transformations of the full unimodular supergravity theory.

Let us introduce the action, $S^{(Aux)}$, for the auxiliary fields
\begin{equation}
S^{(Aux)}=-\frac{1}{3}\idx\,\big(S^2+P^2+A^{\mu}A_{\mu}\big)
\label{SAUX}
\end{equation}

It can be readily shown that
\begin{equation*}
\begin{array}{l}
{\delta^{L}_{\epsilon}S^{(LUG)}=\frac{i}{4}\idx\,\overline{\epsilon}\gamma^{\mu}\psi^{\nu}\big[\partial^2 h_{\mu\nu}+\partial_{\mu}\partial_{\rho}h^{\rho}_{\nu}+\partial_{\nu}\partial_{\rho}h^{\rho}_{\mu}\big],}\\[4pt]
{\delta^{L}_{\epsilon}S^{(RS)}=-\frac{i}{4}\idx\,\overline{\epsilon}\gamma^{\mu}\psi^{\nu}\big[\partial^2 h_{\mu\nu}-\partial_{\mu}\partial_{\rho}h^{\rho}_{\nu}-\partial_{\nu}\partial_{\rho}h^{\rho}_{\mu}\big]}\\[4pt]
{\phantom{\delta^{L}_{\epsilon}S^{(RS)}=\,\,\,}+\idx\, \big[\frac{1}{3}\overline{\epsilon}
(S-i\gamma_5 P)\partial^{\mu}\psi_{\mu}+\frac{1}{2}\overline{\epsilon}\gamma_5\gamma^{\rho\sigma\lambda}\partial_{\rho}\psi_{\lambda}A_{\rho}+\frac{1}{3}\overline{\epsilon}\gamma_5
\Aslash\,\partial^{\lambda}\psi_{\lambda}
}\big]\\[4pt]
{\delta^{L}_{\epsilon}S^{(Aux)}=-\idx\, \big[\frac{1}{3}\overline{\epsilon}(S-i\gamma_5 P)\partial^{\mu}\psi_{\mu}+\frac{1}{2}\overline{\epsilon}\gamma_5\gamma^{\rho\sigma\lambda}\partial_{\rho}\psi_{\lambda}A_{\rho}
+\frac{1}{3}\overline{\epsilon}\gamma_5
\Aslash\,\partial^{\lambda}\psi_{\lambda}}\big]
\end{array}
\end{equation*}
Hence, if we define
\begin{equation}
S^{(LUSG)}=S^{(LUG)}+S^{(RS)}+S^{(Aux)},
\label{LUSGAction}
\end{equation}
we get
\begin{equation*}
\delta^{L}_{\epsilon}S^{(LUSG)}=0.
\end{equation*}
We are thus entitled to define $S^{(LUSG)}$ as the action of the off-shell linearized unimodular $N=1, d=4$ supergravity theory.

Our next task will be the computation of the commutator $[\delta^L_{\epsilon_1},\delta^L_{\epsilon_2}]$. Before  carrying out such computation we shall establish  the following variations of arbitrary --and therefore not restricted by the unimodular constraints in (\ref{tracelessness}) and (\ref{RSgauge})-- $h^{\mu}_{\mu}$ and $\psi_{\mu}$
\begin{equation}
\delta^L_{\epsilon_1}\delta^L_{\epsilon_2}\big[h^{\mu}_{\mu}\big]|_{[h=0,\gamma\cdot \psi=0]}=0,\quad \delta^L_{\epsilon_1}\delta^L_{\epsilon_2}\big[\gamma^{\mu}\psi_{\mu}\big]|_{[h=0,\gamma\cdot \psi=0]}=0.
\label{keyresult}
\end{equation}
The symbol $|_{[h=0,\gamma\cdot \psi=0]}$ indicates that  the unimodular constraints $h\equiv h^{\mu}_{\mu}=0$ and $\gamma\cdot\psi\equiv\gamma^{\mu}\psi_{\mu}=0$ are imposed after having worked out the infinitesimal variations. The action of $\delta^L_{\epsilon}$ on the arbitrary fields is given by the definitions in (\ref{susytrans}) and (\ref{compensationgaugetrans}) with the unimodular constraints removed.

Now, because $\theta(x)$ in (\ref{compensationgaugetrans}) makes sense for arbitrary fields, it is plane that the following equation
\begin{equation*}
\epsilon_{2}\Psi[h_{\mu\nu},\psi_{\mu},S,P,A_{\mu}]\equiv\delta^L_{\epsilon_2}\big[\gamma^{\mu}\psi_{\mu}\big]=0,
\end{equation*}
holds for arbitrary fields --ie, not constrained by $h=0$ and $\gamma\cdot\psi=0$,  by construction. Hence, its variation under $\delta^L_{\epsilon_1}$ also vanishes. Hence,
\begin{equation}
\delta^L_{\epsilon_1}\delta^L_{\epsilon_2}\big[h^{\mu}_{\mu}\big]=\epsilon_2\delta^L_{\epsilon_1}\big[\gamma^{\mu}\psi_{\mu}\big]=0
\label{doublevariation}
\end{equation}

We are now ready to work out the action of $[\delta^L_{\epsilon_1},\delta^L_{\epsilon_2}]$ on the fields. Due to the results  (\ref{vanishingvarlin}) and (\ref{keyresult}), one can readily do so by computing the   action  of  such commutators on arbitrary --ie, not constrained by $h=0$ and $\gamma\cdot\psi=0$-- fields and then imposing the unimodular constraints on the result. Now,  notice that if we remove the summand  $\partial_{\mu}\theta$ from  the transformations in (\ref{susytrans}), we are left with the standard linearized off-shell supergravity transformations, whose algebra closes on translations --modulo gauge  transformations when the commutator acts of either $h_{\mu\nu}$ or
$\psi_{\mu}$. Hence, it is not difficult to reach the conclusion that the following equations hold for the fields --constrained-- of our linearized unimodular supergravity theory:
\begin{equation}
\begin{array}{l}
{[\delta^L_{\epsilon_1},\delta^L_{\epsilon_2}]h_{\mu\nu}=\frac{i}{2}\overline{\epsilon}_1\gamma^{\rho}\epsilon_2\,\partial_{\rho}h_{\mu\nu}+
\partial_{\mu}\chi_{\nu}+\partial_{\nu}\chi_{\mu},}\\[4pt]
{[\delta^L_{\epsilon_1},\delta^L_{\epsilon_2}]\psi_{\mu}=\frac{i}{2}\overline{\epsilon}_1\gamma^{\rho}\epsilon_2\,\partial_{\rho}\psi_\mu}{+\partial_{\mu}
\Theta,}\\[4pt]
{[\delta^L_{\epsilon_1},\delta^L_{\epsilon_2}]S=\frac{i}{2}\overline{\epsilon}_1\gamma^{\rho}\epsilon_2\,\partial_{\rho}S,}\\[4pt]
{[\delta^L_{\epsilon_1},\delta^L_{\epsilon_2}]P=\frac{i}{2}\overline{\epsilon}_1\gamma^{\rho}\epsilon_2\,\partial_{\rho}P,}\\[4pt]
{[\delta^L_{\epsilon_1},\delta^L_{\epsilon_2}]A_{\mu}=\frac{i}{2}\overline{\epsilon}_1\gamma^{\rho}\epsilon_2\,\partial_{\rho}A_{\mu},}
\end{array}
\label{susyalgebra}
\end{equation}
with
\begin{equation*}
\begin{array}{l}
{\chi_{\nu}=-\frac{i}{4}\overline{\epsilon}_1\gamma^{\mu}\epsilon_2 h_{\mu\nu}+\frac{i}{2}\big(\overline{\epsilon}_1\gamma_{\nu}\theta[\epsilon_2]-
\overline{\epsilon}_2\gamma_{\nu}\theta[\epsilon_1]\big)}\\[4pt]
{\Theta=-\frac{i}{2}\overline{\epsilon}_1\gamma^{\rho}\epsilon_2 \psi_{\rho}+
\frac{i}{8}(\overline{\epsilon}_1\gamma^{\sigma}\psi^{\rho}\gamma_{\rho\sigma}\epsilon_2 -\overline{\epsilon}_2\gamma^{\sigma}\psi^{\rho}\gamma_{\rho\sigma}\epsilon_1)+
\delta^L_{\epsilon_1}\theta[\epsilon_2]-\delta^L_{\epsilon_2}\theta[\epsilon_1]}.
\end{array}
\end{equation*}
$\theta[\epsilon]$ is given in (\ref{compensationgaugetrans}).

Using the value of $\theta[\epsilon]$  given in (\ref{compensationgaugetrans}), one can show that this $\chi_{\mu}$ satisfies
\begin{equation*}
\partial_{\mu}\chi^{\mu}=0,
\end{equation*}
if $h^{\mu}_{\mu}=0$. Hence, $\chi_{\mu}$ defines an infinitesimal transverse diffeomorphism, as required. Further, $\gamma^\mu\psi_\mu=0$, (\ref{doublevariation}) and (\ref{susyalgebra}) leads to the conclusion
that
$$
\prslash\, \Theta =0.
$$
Hence, $\Theta$ defines a gauge transformation, $\partial_\mu \Theta$, of $\psi_\mu$ which preserves the constraint $\gamma^\mu\psi_\mu=0$.

It is clear that the algebra generated by the transformations in (\ref{susyalgebra}) closes on translations when these transformations act on local operators which are invariant under the gauge transformations in (\ref{Tdiff}) and ({\ref{gtrans}). This invariance being a sensible requirement for a local operator to qualify as an observable. We thus conclude that the fields of the
linearized supergravity theory with action in (\ref{LUSGAction}) carry a linear representation of the $N=1$ supersymmetry algebra in four dimensions.

 Let us now focus on the plane wave solutions to the equations of motion derived from $S^{(RS)}$ in (\ref{RSaction}) with
 $\psi_{\mu}$ such that $\gamma^{\mu}\psi_{\mu}=0$. We shall close this section by showing that such solutions involve only helicity $\pm 3/2$ quanta upon canonical quantization.

 By setting to zero the change of $S^{(RS)}$ under the variation
 \begin{equation*}
 \delta\psi_{\mu}=\big(\delta_{\mu}^{\nu}-\frac{1}{4}\gamma_{\mu}\gamma^{\nu}\big)\delta\sigma_{\nu},
 \end{equation*}
 where $\delta\sigma_{\nu}$ is an  arbitrary infinitesimal --ie, not constrained-- spinor-vector field, one obtains the equation of motion to be satisfied by the $\psi_{\mu}$ of our linearized unimodular  supergravity theory. This equation reads
 \begin{equation*}
 \big(\delta^{\mu}_{\nu}-\frac{1}{4}\gamma^{\mu}\gamma_{\nu}\big)\gamma^{\nu\rho\sigma}\partial_{\rho}\psi_{\sigma}=0.
 \end{equation*}
Since $\gamma^{\mu}\psi_{\mu}=0$, the previous equation of motion is equivalent to
\begin{equation}
\prslash \psi_{\mu}=\frac{1}{2}\gamma_{\mu}\big(\partial^{\rho}\psi_{\rho}\big).
\label{ugravitinoeq}
\end{equation}
We shall use below the fact that the previous equation implies the following one
\begin{equation}
\prslash\big(\partial^{\mu}\psi_{\mu}\big)=0.
\label{impliedeq}
\end{equation}

A general Majorana plane-wave solution to the previous equation has following  form in terms of its positive and negative frequency parts:
\begin{equation}
\psi_{\mu}(x)\,=\, \id3k\;\big[\psi^{(+)}_{\mu}(\vec{k})\,e^{ikx}\,+\,\psi^{(-)}_{\mu}(\vec{k})\,e^{-ikx}\big]
\label{planewave}
\end{equation}
with $E(\vec{k})=\vec{k}^2$, $kx=E(\vec{k})x^0-\vec{k}\cdot\vec{x}$, $\psi^{(+)}_{\mu}=C[\overline{\psi}^{(-)}_{\mu}]^{\top}$ and $\gamma^{\mu}\psi^{(\pm)}_{\mu}=0$.

By substituting (\ref{planewave}) in (\ref{ugravitinoeq}), one gets
\begin{equation}
\kslash \psi_{\mu}^{(\pm)}(\vec{k})=\frac{1}{2}\gamma_{\mu}\big(k^{\rho}\psi^{(\pm)}_{\rho}(\vec{k})\big),
\label{mastereq}
\end{equation}
where $k^{\rho}=(E(\vec{k}),\vec{k})$. Obviously, (\ref{impliedeq}) leads to
\begin{equation}
\kslash\big(k^{\rho}\psi^{(\pm)}_{\rho}(\vec{k})\big)=0.
\label{muchneeded}
\end{equation}

Now, multiplying both sides of (\ref{mastereq}) by $\kslash$, first, and then taking into account that $k^2=0$, that $\gamma^{\mu}\psi^{(\pm)}_{\mu}=0$ and that  (\ref{muchneeded}) holds, one reaches the conclusion  that
\begin{equation*}
k^{\mu}\psi^{(\pm)}_{\mu}(\vec{k})=0.
\end{equation*}

Putting it all together, we conclude that the $\psi^{\pm}_{\mu}(\vec{k})$'s of our plane-wave function in (\ref{planewave}) have to satisfy the following equations
\begin{equation*}
\kslash\, \psi^{(\pm)}_{\mu}(\vec{k})=0, \quad k^{\mu}\psi^{(\pm)}_{\mu}(\vec{k})=0 \quad\text{and}\quad \gamma^{\mu}\psi^{(\pm)}_{\mu}(\vec{k})=0.
\end{equation*}

It is clear that the solution to the previous set of equations contains  longitudinal modes of the type $k_{\mu}\phi^{\pm}(\vec{k})$, $\phi^{\pm}(\vec{k})$ being spinors that satisfy $\kslash\,\phi^{\pm}(\vec{k})=0$.   And yet, this longitudinal modes can be gauged away while preserving the constraint $\gamma^{\mu}\psi_{\mu}=0$, for $\kslash\,\phi^{\pm}(\vec{k})=0$. Finally, it is a well-established fact --see, eg, \cite{VanNieuwenhuizen:1981ae}-- that once these longitudinal modes are disposed of, we are left  only with modes which, upon quantization, give  rise  to operators whose  helicity is either $+3/2$ or $-3/2$.

To close this section we shall compute the supersymmetry current that the  Noether's theorem associates to the supersymmetry transformations in (\ref{susytrans}). Let replace the rigid parameter $\epsilon$ in (\ref{susytrans}) with
an $x$-dependent infinitesinal Majorana spinor $\epsilon(x)$ and define the following transformations
\begin{equation}
\begin{array}{l}
{\delta^{L}_{\epsilon(x)} h_{\mu\nu}\,=\,-\frac{i}{2}\,\overline{\epsilon}(x)(\gamma_{\mu}\psi_{\nu}\,+\,\gamma_{\nu}\psi_{\mu}),}\\[4pt]
{\delta^{L}_{\epsilon(x)} \psi_{\mu}\,=\,-\frac{1}{4}\,\partial^{\rho}h^{\sigma}_\mu\,\gamma_{\rho\sigma}\epsilon(x)+\frac{i}{6}\gamma_{\mu}(S-i\gamma_5 P)\epsilon(x)+\frac{i}{2}\gamma_5 (A_{\mu}-\frac{1}{3}\gamma_{\mu}\gamma^{\nu}A_{\nu})\epsilon(x)+\partial_{\mu}\theta[\epsilon(y)],}\\[4pt]
{\delta^{L}_{\epsilon(x)} S\,=\,-\frac{1}{2}\,\overline{\epsilon}(x)\gamma^{\rho\sigma}\partial_{\rho}\psi_{\sigma},}\\[4pt]
{\delta^{L}_{\epsilon(x)} P\,=\,\frac{i}{2}\,\overline{\epsilon}(x)\gamma_{5}\gamma^{\rho\sigma}\partial_{\rho}\psi_{\sigma},}\\[4pt]
{\delta^{L}_{\epsilon(x)} A_{\mu}\,=\,\frac{3}{4}\overline{\epsilon}(x)\gamma_{5}(\eta_{\mu\rho}-\frac{1}{3}\gamma_{\mu}\gamma_{\rho})
\gamma^{\rho\sigma\lambda}\partial_{\sigma}\psi_{\lambda},}
\end{array}
\label{susytranslocal}
\end{equation}
\begin{equation}
\begin{array}{l}
{\theta[\epsilon(y)](x)=-\idy\,D(x-y)\big[-\frac{1}{4}\,\gamma^\mu\partial^{\rho}h^{\sigma}_\mu(y)\,\gamma_{\rho\sigma}\epsilon(y)+\frac{2i}{3}(S(y)-i\gamma_5 P(y))\epsilon(y)+\frac{i}{6}\gamma_5 \gamma^{\mu}A_{\mu}(y)\epsilon(y)\big],}\\[4pt]
{\prslash_{x}D(x-y)=\delta(x-y)}\\[4pt]
{\prslash_x \theta[\epsilon(y)](x)=\frac{1}{4}\,\gamma^\mu\partial^{\rho}h^{\sigma}_\mu\,\gamma_{\rho\sigma}\epsilon(x)-\frac{2i}{3}(S-i\gamma_5 P)\epsilon(x)-\frac{i}{6}\gamma_5 \gamma^{\mu}A_{\mu}\epsilon(x)}
\end{array}
\label{compensationgaugetranslocal}
\end{equation}

Now, since $S^{LUG}$ is invariant under the rigid supersymmetry transformations in (\ref{susytrans}), one concludes that
\begin{equation}
\delta^{L}_{\epsilon(x)}S^{(LUSG)} = \idx\,\partial_{\mu}\overline{\epsilon}(x)\, J^{\mu}(x)= -\idx\,\overline{\epsilon}(x)\partial_{\mu}J^{\mu}(x).
\label{localvar}
\end{equation}

Taking into account (\ref{SLUG}), (\ref{RSaction}), (\ref{SAUX}) and (\ref{susytranslocal}), one shows that
\begin{equation}
\delta^{L}_{\epsilon(x)}S^{(LUSG)} = \delta^{L}_{\epsilon(x)}S^{(LUG)}+\delta^{L}_{\epsilon(x)}S^{(RS)}+\delta^{L}_{\epsilon(x)}S^{(Aux)}=
\idx\,\partial_{\mu}\overline{\epsilon}(x)\,\left[\frac{i}{4}\partial^{\rho}h^{\sigma}_{\lambda}\gamma_{\rho\sigma}\gamma^{\lambda\mu\delta}\psi_{\delta}\right].
\label{localvarcomp}
\end{equation}
By comparing (\ref{localvar}) and (\ref{localvarcomp}), one concludes that the supersymmetry current, $J^{\mu}$, associated to supersymmetry transformations in (\ref{susytrans}) reads
\begin{equation}
J^{\mu}=\frac{i}{4}\partial^{\rho}h^{\sigma}_{\lambda}\gamma_{\rho\sigma}\gamma^{\lambda\mu\delta}\psi_{\delta}\big.
\label{supercu}
\end{equation}
That this supersymmetry current is conserved  when $h_{\mu\nu}$ and $\psi_{\mu}$ satisfy the equation of motion derived from $S^{(LUSG)}$ is a consequence of the fact that the variations in (\ref{susytranslocal}) preserve the unimodularity constraints $h^{\mu}_{\mu}=0$ and $\gamma^{\mu}\psi_{\mu}=0$ and, hence,
\begin{equation}
\delta^{L}_{\epsilon(x)}S^{(LUSG)}=0,
\end{equation}
if $h_{\mu\nu}$ and $\psi_{\mu}$ are solutions the equation of motion. Recall that $\epsilon(x)$ in  (\ref{localvar}) is arbitrary Majorana spinor.

Notice that $J^{\mu}$ in (\ref{supercu}) is the very supersymmetry current that one obtains by applying, first, the technique above to the ordinary linearized supergravity theory and, then,  imposing on the fields the gauge conditions $h^{\mu}_{\mu}=0$ and $\gamma^{\mu}\psi_{\mu}=0$. Recall that, with our conventions,  the \textit{Fierz-Pauli} action, $S_{FP}$ reads
\begin{equation}
S^{(LUG)}\,=\,-\frac{1}{4}\idx\,\Big[\frac{1}{2}\,h^{\mu\nu}\partial^2 h^{\mu\nu}\,+\,\partial_{\mu}h^{\mu\lambda}\partial_{\nu}h^{\nu}_{\phantom{\nu}\lambda}+h\partial_{\mu}\partial_{\nu}h^{\mu\nu}-\frac{1}{2}h\partial^2 h\Big],
\end{equation}
where $h=h^{\mu}_{\mu}$.

Finally, the current $J^{\mu}$ in (\ref{supercu}) plays a mayor role in the construction of the interacting unimodular supergravity theory by means of the Noether method \cite{West:1986wua} as discussed in the next section.

\newpage

\section{Off-shell unimodular $N=1,d=4$ supergravity}

In this section we shall introduce the unimodular $N=1,d=4$ supergravity in its off-shell formulation. But, first, let us settle the notation.

$\eta_{ab}$ will denote the Minkowski metric with mostly minus signature. $g_{\mu\nu}$ will stand for the metric of the semi-Riemannian $4d$ spin manifold. $e^a_{\mu}$ denotes a vierbein for the metric $g_{\mu\nu}$ and $e^{\mu}_a$  the inverse of the former. $\omega^{\phantom{\mu}ab}_{\mu}$ will stand for the spin connection and
$R^{ab}_{\mu\nu}[\omega]$ the curvature of the latter:
\begin{equation*}
R^{ab}_{\mu\nu}[\omega]=\partial_{\mu}\omega^{\phantom{\nu}ab}_{\nu}-\partial_{\nu}\omega^{\phantom{\mu}ab}_{\mu}+\omega^{\phantom{\mu}ac}_{\mu}\omega^{\phantom{\nu c}b}_{\nu c}-
\omega^{\phantom{\nu}ac}_{\nu}\omega^{\phantom{\nu c}b}_{\mu c}
\end{equation*}
The numerical Dirac matrices will be denoted by $\gamma^{a}$ and they satisfy
\begin{equation*}
\{\gamma^{a},\gamma^{a}\}\,=\,2\eta^{ab}.
\end{equation*}
The matrix $\gamma^{\mu}$ is defined by the equation $\gamma^{\mu}=\gamma^a e_a^{\mu}$.  $\psi_{\mu}$ will be the symbol representing a Majorana spin-3/2 field on the manifold. $\overline{\psi}_{\mu}=\psi_{\mu}^{\dagger}\gamma^0$ wil denote the Dirac conjugate of $\psi_{\mu}$. $D_{\mu}[\omega^{ab}_\rho]$  will act on
$\psi_{\nu}$ as follows
\begin{equation*}
D_{\mu}[\omega^{ab}_\rho]\psi_{\nu}=\partial_{\mu}\psi_{\nu}+\frac{1}{4}\omega^{\phantom{\mu}ab}_{\mu}\gamma_{ab}\psi_{\nu},
\end{equation*}
where $\gamma_{ab}=\frac{1}{2}[\gamma_a,\gamma_b]$. The following symbol will be much used
\begin{equation*}
\gamma^{\mu_1\mu_2\mu_3}=\frac{1}{3!}\sum_{\pi}\,(-1)^{\sigma_{\pi}}
\,\gamma^{\mu_{\pi(1)}}\gamma^{\mu_{\pi(2)}}\gamma^{\mu_{\pi(3)}}.
\end{equation*}
$\pi(1)\pi(2)\pi(3)$ denotes a permutation of $123$ with signature $\sigma_{\pi}$.

In view of the results presented in the previous section it is quite natural to postulate that the action of the theory at hand should be
\begin{equation}
S^{(USG)}\,=-\frac{1}{2\kappa^2}\!\!\idx\, e^{\mu}_a e^{\nu}_b R^{ab}_{\mu\nu}[\omega(e^c_\rho,\psi_\rho)]-\frac{i}{2}\!\!\idx\,\overline{\psi}_\mu\gamma^{\mu\nu\rho}D_{\nu}[\omega(e^a_\rho,\psi_\rho)]\psi_{\rho}-\frac{1}{3}\!\!\idx
\,\big[S^2+P^2+A^{a}A_{a}.\big]
\label{SUSG}
\end{equation}
In the previous equation the vierbein, $e^{a}_{\mu}$, and the field $\psi_{\mu}$ are constrained by the following unimodularity conditions:
\begin{equation}
e\equiv \rm{det}\,e^a_\mu =1,\quad \gamma^{\mu}\psi_{\mu}=0.
\label{fulluconstraints}
\end{equation}
The constraint on the determinant of $e^{a}_{\mu}$ is what defines \cite{Alvarez:2015oda} the unimodular gravity theory in the Palatini formalism. The invariance of the unimodularity --ie, $e=1$-- of $e^a_{\mu}$  under the supergravity transformations is guaranteed by the constraint on
$\psi_{\mu}$, as we shall see below. $S$, $P$ and $A_{\mu}$ are the auxiliary fields needed to set up the off-shell formalism. $S$ is a real scalar, whereas $P$ and $A_{\mu}$ are a real pseudoscalar and real pseudovector, respectively. In the action in (\ref{SUSG}), $\omega(e^a_\rho,\psi_\rho)$ denotes the following spin connection with torsion:
\begin{equation}
\begin{array}{l}
{\omega^{\phantom{\mu}ab}_{\mu}(e^c_\sigma,\psi_\sigma)=\omega^{\phantom{\mu}ab}_{\mu}(e^c_\sigma)+K^{\phantom{\mu}ab}_{\mu}(e^c_\sigma,\psi_\sigma),}\\[4pt]
{K^{\phantom{\mu}ab}_{\mu}(e^c_\sigma,\psi_\sigma)=i\frac{\kappa^2}{4}\big(\overline{\psi}_{\mu}\gamma^b\psi^a-\overline{\psi}^a\gamma_{\mu}\psi^b
+\overline{\psi}^b\gamma^a\psi_\mu\big),}
\end{array}
\label{spincon}
\end{equation}
where $\omega^{\phantom{\mu}ab}_{\mu}(e^c_\sigma)$ is the Levi-Civita spin connection for the vierbein $e^a_{\mu}$. The reader may notice that $S^{(USG)}$ in (\ref{SUSG}) is the standard action \cite{VanNieuwenhuizen:1981ae} of $N=1, d=4$ supergravity when $e^{a}_\mu$ and $\psi_{\mu}$ satisfy the constraints in (\ref{fulluconstraints}).

To define the supergravity transformations that will leave invariant $S^{(USG)}$ in (\ref{SUSG}), we shall proceed as follows. First, we shall recall the value of the supergravity transformations of standard $N=1, d=4$ supergravity:
\begin{equation}
\begin{array}{l}
{\tilde{\delta}_{\tilde{\epsilon}}\tilde{e}^a_{\mu}=-i\frac{\kappa}{2}\overline{\tilde{\epsilon}}\gamma^a\tilde{\psi}_{\mu},\quad \quad \tilde{\gamma}_{\mu}\equiv \gamma_a \tilde{e}^a_{\mu}}\\[4pt]
{\tilde{\delta}_{\tilde{\epsilon}}\tilde{\psi}_{\mu}=\frac{1}{\kappa}D_{\mu}[\omega(\tilde{e}^a_{\sigma},\tilde{\psi}_{\sigma})]\tilde{\epsilon}
+\frac{i}{6}\tilde{\gamma}_{\mu}(S-i\gamma_5 P)\tilde{\epsilon} +\frac{i}{2}\gamma_5(\delta^{\nu}_{\mu}-\frac{1}{3}\tilde{\gamma}_{\mu}\tilde{\gamma}^{\nu})\tilde{\epsilon}A_{\nu},}\\[4pt]
{\tilde{\delta}_{\tilde{\epsilon}}S=-\frac{1}{4}\overline{\tilde{\epsilon}}\tilde{\gamma}_{\mu}\tilde{{\cal R}}^{\mu},\quad\quad}\\[4pt]
{\tilde{\delta}_{\tilde{\epsilon}}P=\frac{i}{4}\overline{\tilde{\epsilon}}\gamma_5\tilde{\gamma}_{\mu}\tilde{{\cal R}}^{\mu}}\\[4pt]
{\tilde{\delta}_{\tilde{\epsilon}}A^a=\frac{3}{4}\overline{\tilde{\epsilon}}\gamma_5(\tilde{e}^a_{\nu}-\frac{1}{3}\gamma^a\tilde{\gamma}_{\nu})
\tilde{{\cal R}}^{\nu},}
\end{array}
\label{standarsugratrans}
\end{equation}
where $\tilde{e}^a_{\mu}$, $\tilde{\psi}_{\mu}$ are, respectively, the vierbein and gravitino fields of standard supergravity, and, therefore, they are not subjected to the constrains in (\ref{fulluconstraints}), and $S$, $P$ and $A_a$ are the auxiliary fields.
$\tilde{\epsilon}$ is the standard supergravity transformation parameter. $\cal{R}^{\mu}$ is given by the formulae
\begin{equation*}
\begin{array}{l}
{\tilde{{\cal R}}^\mu=\tilde{\gamma}^{\mu\nu\rho}\tilde{{\cal D}}_{\nu}\tilde{\psi}_{\rho},}\\[4pt]
{\tilde{{\cal D}}_{\mu}\tilde{\psi}_{\rho}=D_{\mu}[\omega^{ab}_{\nu}(\tilde{e}^c_{\sigma},\tilde{\psi}_{\sigma})]\tilde{\psi}_{\rho}
-i\frac{\kappa}{6}\tilde{\gamma}_{\rho}(S-i\gamma_5 P)\tilde{\psi}_{\mu} -i\frac{\kappa}{2}\gamma_5(\delta^{\lambda}_{\rho}-\frac{1}{3}\tilde{\gamma}_{\rho}\tilde{\gamma}^{\lambda})\tilde{\psi}_{\mu}A_{\lambda},}\\[4pt]
{D_{\mu}[\omega^{ab}_{\nu}(\tilde{e}^c_{\sigma},\tilde{\psi}_{\sigma})]=\partial_{\mu}+
\frac{1}{4}\omega(\tilde{e}^c_{\sigma},\tilde{\psi}_{\sigma})^{\phantom{\mu}ab}_{\mu}\gamma_{ab}.}
\end{array}
\end{equation*}
$\omega^{ab}_{\nu}(\tilde{e}^c_{\sigma},\tilde{\psi}_{\sigma})$ is the spin connection with torsion of standard $N=1, d=4$ supergravity. This spin connection yields the connection in (\ref{spincon}), when $\tilde{e}^a_\mu=e^a_{\mu}$ and $\tilde{\psi}_{\mu}=\psi_{\mu}$:
\begin{equation*}
\omega^{ab}_{\nu}(e^c_{\sigma},\psi_{\sigma})\,=\,\omega^{ab}_{\nu}(\tilde{e}^c_{\sigma},\tilde{\psi}_{\sigma})\vert_{[\tilde{e}^a_\mu=e^a_\mu, \tilde{\psi}_\mu=\psi_\mu]}.
\end{equation*}
The transformations in (\ref{standarsugratrans}) were introduced by the authors of Ref. \cite{Ferrara:1978em}, but the reader should be warned that our conventions are not theirs.

Taking into account the way the action $S^{(USG)}$ in (\ref{SUSG}) was obtained,
it is quite natural to define the supergravity transformations of the fields in it by setting $\tilde{e}^a_\mu=e^a_{\mu}$ and $\tilde{\psi}_{\mu}=\psi_{\mu}$ in the transformations in (\ref{standarsugratrans}). However this is not enough to obtain a set of meaningful transformations, for it is plain that they will not preserve the constraint $\gamma^{\mu}\psi_{\mu}=0$, if $\tilde{\epsilon}$ is arbitrary. We are thus lead to restrict the set of allowed values of $\tilde{\epsilon}$ to those belonging to the set of solutions of the equation:
\begin{equation}
\delta_{\tilde{\epsilon}}\big(\gamma^{\mu}\psi_{\mu}\big)=
\gamma^{a}\delta_{\tilde{\epsilon}}e_a^{\mu}\psi_\mu+\gamma^{\mu}\delta_{\tilde{\epsilon}}\psi_{\mu}=0,
\label{preservingRSG}
\end{equation}
where $\delta_{\tilde{\epsilon}}e_a^{\mu}$ and $\delta_{\tilde{\epsilon}}\psi_{\mu}$ are defined as follows
\begin{equation*}
\delta_{\tilde{\epsilon}}e_a^{\mu}= \big[\tilde{\delta}_{\tilde{\epsilon}}\tilde{e}^a_{\mu}\big]_{[\tilde{e}^a_\mu=e^a_\mu, \tilde{\psi}_\mu=\psi_\mu]},\quad
\delta_{\tilde{\epsilon}}\psi=\big[\tilde{\delta}_{\tilde{\epsilon}}\tilde{\psi}_{\mu}\big]_{[\tilde{e}^a_\mu=e^a_\mu,\tilde{\psi}_\mu=\psi_\mu]}.
\end{equation*}
The symbols on the right hand sides of the previous equations indicate that $\tilde{e}^a_\mu=e^a_{\mu}$ and $\tilde{\psi}_{\mu}=\psi_{\mu}$ are imposed on the right hand sides of the corresponding  transformations in (\ref{standarsugratrans}). By using the definitions in  (\ref{standarsugratrans}), one easily shows that (\ref{preservingRSG}) is equivalent to
\begin{equation}
\gamma^{\mu}D_{\mu}[\omega(e^a_{\sigma},\psi_{\sigma})]\tilde{\epsilon}+i\frac{\kappa^2}{2} (\overline{\tilde{\epsilon}}\gamma^b\psi_a)\gamma^a\psi_b
+i\frac{2\kappa}{3}(S-i\gamma_5 P)\tilde{\epsilon} +i\frac{\kappa}{6}\gamma_5\gamma^{\nu}\tilde{\epsilon}A_{\nu}=0.
\label{preservingRSGdevelop}
\end{equation}
From now on we shall denote by $\epsilon(e^a_\mu,\psi,S,P,A_\mu)$ --or, just $\epsilon$, for short-- any solution to the equation in (\ref{preservingRSGdevelop}).
We shall take the  solution to (\ref{preservingRSGdevelop}) to be given by the formal series expansion in $\kappa$ that solves the equation upon setting $e^a_\mu=\delta^a_\mu+\sum_{n>1}\kappa^n C^{(n)\,a}_\mu$ --the $C^{(n)\,a}_\mu$'s are constrained by $\det e^a_\mu=1$. Since this formal series expansion can be worked out by sequentially solving an infinite set of inhomogeneous Dirac equations   in flat spacetime, it is plain  that (\ref{preservingRSGdevelop}) imposes on the fields $e^a_\mu$ and $\psi_\mu$ no constraints other than the appropriate regularity and boundary conditions for the solutions to those Dirac equations to be smooth enough. It is in this sense that (\ref{preservingRSGdevelop}) holds whatever the value of $e^a_\mu,\psi_\mu, S, P$ and $A_a$ and, in particular, for fields that differ infinitesimally. To be more concrete, let us work out the first order in $\kappa$ solution to (\ref{preservingRSGdevelop}). Actually, the first order in $\kappa$ contribution to $\epsilon$, gives rise precisely to supersymmetry transformations (\ref{susytrans}), as we had anticipated in the previous section. Indeed, if we expand the metric around the Minkowski metric --$ie, g_{\mu\nu}=\eta_{\mu\nu}+\kappa h_{\mu\nu}+o(\kappa^2)$, the spin connection $\omega^{ab}_{\mu}(e^{c}_\sigma,\psi_\sigma)$ inherits the following expansion in $\kappa$:
\begin{equation}
\omega^{ab}_{\mu}(e^{c}_\sigma,\psi_\sigma)= -\frac{\kappa}{2}(\partial^a h^b_\mu-\partial^b h^a_\mu)+o(\kappa^2).
\label{connectionfirst}
\end{equation}
If we substitute now this expression and $\epsilon= \epsilon^{(0)}+\kappa\epsilon^{(1)} + o(\kappa^2)$ in (\ref{preservingRSGdevelop}), we obtain
\begin{equation}
\begin{array}{l}
{\prslash \epsilon^{(0)}=0}\\[4pt]
{\prslash \epsilon^{(1)}=\frac{1}{4}\,\gamma^\mu\partial^{\rho}h^{\sigma}_\mu\,\gamma_{\rho\sigma}\epsilon^{(0)}-\frac{2i}{3}(S-i\gamma_5 P)\epsilon^{(0)}-\frac{i}{6}\gamma_5 \gamma^{\mu}A_{\mu}\epsilon^{(0)}}.
\end{array}
\label{epsilonexpan}
\end{equation}
Hence, $\epsilon^{(1)}$ can be taken to be given by
\begin{equation}
\begin{array}{l}
{\epsilon^{(1)}=\varepsilon^{(1)}+\idy\,D(x-y)\big[\frac{1}{4}\,\gamma^\mu\partial^{\rho}h^{\sigma}_\mu(y)\,\gamma_{\rho\sigma}\epsilon^{(0)}-\frac{2i}{3}(S(y)-i\gamma_5 P(y))\epsilon^{(0)}-\frac{i}{6}\gamma_5 \gamma^{\mu}A_{\mu}(y)\epsilon^{(0)}\big],}\\[4pt]
{\prslash_{x}D(x-y)=\delta(x-y).}
\end{array}
\label{epsiloone}
\end{equation}
where $\prslash \varepsilon^{(1)}=0$
By choosing a rigid $\epsilon^{(0)}$ and  $\varepsilon^{(1)}=0$, taking into account (\ref{connectionfirst}), (\ref{epsilonexpan}) and (\ref{epsiloone}), and, using, finally,  (\ref{standarsugratrans}), one easily recovers the supersymmetry transformations in (\ref{susytrans}). The full unimodular formalism we are developing is thus naturally , in harmony with the linear unimodular supergravity theory we constructed in the previous section. Note that, indeed, the second equation in (\ref{epsilonexpan}) posses no contraints on the fields, as we said above, for (\ref{epsilonexpan}) always exist provided appropriate regularity and boundary conditions are met.

Notice that the expansion in $\kappa$ we have introduced in the previous paragraph is totally in harmony with the fact that one may rightly consider unimodular supergravity as a theory of gravitons and gravitinos propagating in Minkowski spacetime. Indeed, as we shall see in the next section, the maximally supersymmetric solution  to the equations of motion of the unimodular supergravity theory  is Minkowski spacetime.

Let us briefly discuss  the construction of the full unimodular supergravity theory by using the Noether method --see Reference \cite{West:1986wua}, for the ordinary case. We shall consider the  expansion of the unimodular supergravity action up to first order in $\kappa$. Taking into account (\ref{localvar}) and (\ref{susytranslocal}), we define the following action
\begin{equation}
S_1= S^{(LUSG)}-\frac{\kappa}{2}\idx\; \bar{\psi}_{\mu} J^{\mu}\,+\,o(\kappa^2),
\end{equation}
where $J^{\mu}$ is given in (\ref{supercu}). Let us stress the fact that the previous action can be obtained by expanding  $S^{(USG)}$ in  (\ref{SUSG}) up to first order in $\kappa$.

Now, $S_1$ is invariant, up to order $\kappa$, under the following local transformations
\begin{equation}
\begin{array}{l}
{\delta_{\epsilon(x)} h_{\mu\nu}\,=\,-\frac{i}{2}\,\overline{\epsilon}(x)(\gamma_{\mu}\psi_{\nu}\,+\,\gamma_{\nu}\psi_{\mu}),}\\[4pt]
{\delta_{\epsilon(x)} \psi_{\mu}\,=\frac{1}{\kappa}\partial_{\mu}\epsilon(x)\,-\frac{1}{4}\,\partial^{\rho}h^{\sigma}_\mu\,\gamma_{\rho\sigma}\epsilon(x)+\frac{i}{6}\gamma_{\mu}(S-i\gamma_5 P)\epsilon(x)+\frac{i}{2}\gamma_5 (A_{\mu}-\frac{1}{3}\gamma_{\mu}\gamma^{\nu}A_{\nu})\epsilon(x)+\partial_{\mu}\theta[\epsilon(y)],}\\[4pt]
{\delta_{\epsilon(x)} S\,=\,-\frac{1}{2}\,\overline{\epsilon}(x)\gamma^{\rho\sigma}\partial_{\rho}\psi_{\sigma},}\\[4pt]
{\delta_{\epsilon(x)} P\,=\,\frac{i}{2}\,\overline{\epsilon}(x)\gamma_{5}\gamma^{\rho\sigma}\partial_{\rho}\psi_{\sigma},}\\[4pt]
{\delta_{\epsilon(x)} A_{\mu}\,=\,\frac{3}{4}\overline{\epsilon}(x)\gamma_{5}(\eta_{\mu\rho}-\frac{1}{3}\gamma_{\mu}\gamma_{\rho})
\gamma^{\rho\sigma\lambda}\partial_{\sigma}\psi_{\lambda},}
\end{array}
\label{sugratransnoether}
\end{equation}
But we should also demand that the constraint $\gamma^{\mu}\psi_{\mu}=0$, with $\gamma^a e^\mu_a$, be preserved by the transformations in (\ref{sugratransnoether}) up to order $\kappa^0$ --the transformation for
$\psi_\mu$ starts with $\kappa^{-1}$. Clearly, this will constraint the allowed values of $\epsilon(x)$ in the transformations in (\ref{sugratransnoether}). Using the expansions $e^{\mu}_a=\delta^{\mu}_a-\frac{\kappa}{2} h^\mu_a+o(\kappa^2)$ and $\epsilon(x)=\epsilon^{(0)}+\kappa \varepsilon^{(1)}+o(\kappa^2)$, one gets that $\gamma^{\mu}\psi_{\mu}=0$ is preserved up to order $\kappa^0$, if
\begin{equation*}
\prslash \epsilon^{(0)}=0=\prslash \varepsilon^{(1)}.
\end{equation*}
Bearing in mind this last result and substituting $\epsilon(x)=\epsilon^{(0)}+\kappa \varepsilon^{(1)}+o(\kappa^2)$ in (\ref{sugratransnoether}), one obtains the same transformations rules that are obtained from
(\ref{standarsugratrans}) by expanding in powers of $\kappa$, once $\tilde{\epsilon}$ is constrained by (\ref{preservingRSGdevelop}), ie, once (\ref{epsilonexpan}) and (\ref{epsiloone}) are imposed.

It is high time that we postulate what the unimodular supergravity transformations, $\delta_{\epsilon}e^a_{\mu},\;  \delta_{\epsilon}\psi_{\mu},\;\delta_{\epsilon}S,\;\delta_{\epsilon}P$ and $\delta_{\epsilon}A_a, $ are. This we do now:
\begin{equation}
\begin{array}{l}
{\delta_{\epsilon}e^a_{\mu}=[\tilde{\delta}_{\tilde{\epsilon}}\tilde{e}^a_{\mu}]_{[\tilde{\epsilon}=\epsilon,\tilde{e}^a_{\mu}=e^a_{\mu},
\tilde{\psi}_{\mu}=\psi_{\mu}]}
=-i\frac{\kappa}{2}\overline{\epsilon}\gamma^a\psi_{\mu},\quad \quad \gamma_{\mu}\equiv \gamma_a e^a_{\mu}}\\[4pt]
{\delta_{\epsilon}\psi_{\mu}=[\tilde{\delta}_{\tilde{\epsilon}}\tilde{\psi}_{\mu}]_{[\tilde{\epsilon}=\epsilon,\tilde{e}^a_{\mu}=e^a_{\mu},
\tilde{\psi}_{\mu}=\psi_{\mu}]}
=\frac{1}{\kappa}D_{\mu}[\omega^{ab}_\mu(e^c_{\sigma},\psi_{\sigma})]\epsilon
+\frac{i}{6}\gamma_{\mu}(S-i\gamma_5 P)\epsilon +\frac{i}{2}\gamma_5(\delta^{\nu}_{\mu}-\frac{1}{3}\gamma_{\mu}\gamma^{\nu})\epsilon  A_{\nu},}\\[4pt]
{\delta_{\epsilon}S=[\tilde{\delta}_{\tilde{\epsilon}}S]_{[\tilde{\epsilon}=\epsilon,\tilde{e}^a_{\mu}=e^a_{\mu},
\tilde{\psi}_{\mu}=\psi_{\mu}]}=-\frac{1}{4}\overline{\epsilon}\gamma_{\mu}{\cal R}^{\mu},\quad\quad}\\[4pt]
{\delta_{\epsilon}P=[\tilde{\delta}_{\tilde{\epsilon}}P]_{[\tilde{\epsilon}=\epsilon,\tilde{e}^a_{\mu}=e^a_{\mu},
\tilde{\psi}_{\mu}=\psi_{\mu}]}=\frac{i}{4}\overline{\epsilon}\gamma_5\gamma_{\mu}{\cal R}^{\mu}}\\[4pt]
{\delta_{\epsilon}A^a=[\tilde{\delta}_{\tilde{\epsilon}}A^a]_{[\tilde{\epsilon}=\epsilon,\tilde{e}^a_{\mu}=e^a_{\mu},
\tilde{\psi}_{\mu}=\psi_{\mu}]}=\frac{3}{4}\overline{\epsilon}\gamma_5(\tilde{e}^a_{\nu}-\frac{1}{3}\gamma^a\gamma_{\nu})
{\cal R}^{\nu},}
\end{array}
\label{unisugratrans}
\end{equation}
where
\begin{equation*}
\begin{array}{l}
{{\cal R}^\mu=\gamma^{\mu\nu\rho}{\cal D}_{\nu}\psi_{\rho},}\\[4pt]
{{\cal D}_{\mu}\psi_{\rho}=D_{\mu}[\omega^{ab}_{\nu}(e^c_{\sigma},\psi_{\sigma})]\psi_{\rho}
-i\frac{\kappa}{6}\gamma_{\rho}(S-i\gamma_5 P)\psi_{\mu} -i\frac{\kappa}{2}\gamma_5(\delta^{\lambda}_{\rho}-\frac{1}{3}\gamma_{\rho}\gamma^{\lambda})\psi_{\mu}A_{\lambda},}\\[4pt]
{D_{\mu}[\omega^{ab}_{\nu}(e^c_{\sigma},\psi_{\sigma})]=\partial_{\mu}+
\frac{1}{4}\omega(e^c_{\sigma},\psi_{\sigma})^{\phantom{\mu}ab}_{\mu}\gamma_{ab}.}
\end{array}
\end{equation*}
$\omega^{ab}_{\nu}(e^c_{\sigma},\psi_{\sigma})$ is the spin connection with torsion in (\ref{spincon}). It is most important to recall that $\epsilon$ is a solution to (\ref{preservingRSGdevelop}), so that $\delta_{\epsilon}\big(\gamma^\mu\psi_\mu\big)=0$. The symbol
\begin{equation*}
[\tilde{\delta}_{\tilde{\epsilon}}(\text{field})]_{[\tilde{\epsilon}=\epsilon,\tilde{e}^a_{\mu}=e^a_{\mu},
\tilde{\psi}_{\mu}=\psi_{\mu}]},\quad \text{field}=\tilde{e}^a_{\mu},\;\tilde{\psi}_{\mu},\;S,\;P,\;A_a
\end{equation*}
indicates that the substitutions $\tilde{\epsilon}\rightarrow\epsilon$, $\tilde{e}^a_{\mu}\rightarrow e^a_{\mu}$ and
$\tilde{\psi}_{\mu}\rightarrow\psi_{\mu}$ are applied to the polynomial in  the fields and  their derivatives that are equal to the symbol
$\tilde{\delta}_{\tilde{\epsilon}}(\text{field})$ according to the definition in (\ref{standarsugratrans}).

Taking into account that
\begin{equation*}
\begin{array}{l}
{\delta_{\epsilon}e=[\tilde{\delta}_{\tilde{\epsilon}}\tilde{e}]_{[\tilde{\epsilon}=\epsilon,\tilde{e}^a_{\mu}=e^a_{\mu},
\tilde{\psi}_{\mu}=\psi_{\mu}]}=0,\quad \text{where}\quad e=\det\,e^a_\mu,\quad \tilde{e}=\det\,\tilde{e}^a_\mu,}\\[4pt]
{\delta_{\epsilon}\big(\gamma^\mu\psi_\mu\big)=
[\tilde{\delta}_{\tilde{\epsilon}}\big(\tilde{\gamma}^\mu\tilde{\psi}_\mu\big)]_{[\tilde{\epsilon}=\epsilon,\tilde{e}^a_{\mu}=e^a_{\mu},
\tilde{\psi}_{\mu}=\psi_{\mu}]}=0,}
\end{array}
\end{equation*}
and the definitions in (\ref{unisugratrans}), one concludes that
\begin{equation}
\delta_{\epsilon}\,S^{(USG)}=0,
\label{vanishSUSG}
\end{equation}
where $S^{(USG)}$ is the off-shell unimodular supergravity action in (\ref{SUSG}). Indeed, if we consider the spin connection,  $\omega^{ab}_\mu$, in (\ref{SUSG}) to be an independent field, its equation of motion is solved by $\omega^{\phantom{\mu}ab}_{\mu}(e^c_\sigma,\psi_\sigma)$ in (\ref{spincon}), so that one may apply  the ''1.5'' formalism --see Reference \cite{VanNieuwenhuizen:1981ae}-- to the case at hand. Thus, one obtains

\begin{equation*}
\delta_{\epsilon}\,S^{(USG)}=\delta S_{EH}+\delta S_{RS}+\delta S_{Aux}=0,
\end{equation*}
for
\begin{equation*}
\begin{array}{l}
{\delta S_{EH}=-\frac{1}{2\kappa^2}\!\!\idx\, (\delta_{\epsilon}e^{\mu}_a) e^{\nu}_b R^{ab}_{\mu\nu}[\omega(e^c_\a,\psi_\a)]=-\frac{i}{2\kappa}\!\!\idx\,(\overline{\epsilon}\gamma^\m\psi_a ) e^{\nu}_b R^{ab}_{\m\n}[\omega(e^c_\a,\psi_\a)] ,}\\[4pt]
{\delta S_{RS}=
	\frac{1}{2}\!\!\idx\,\epsilon^{\mu\nu\rho\sigma}[(\delta_{\epsilon}\overline{\psi}_\m\gamma_\sigma\gamma_5 D_{\nu}[\omega(e^c_\a,\psi_\a)]\psi_{\rho}+\overline{\psi}_\m\gamma_a(\delta_{\epsilon}e_{\sigma}^a)\gamma_5 D_{\nu}[\omega(e^c_\a,\psi_\a)]\psi_{\rho}}\\[4pt]
{\phantom{\delta S_{RS}=+\frac{1}{2}\!\!\idx\,\epsilon^{\mu\nu\rho\sigma}[}
	+\overline{\psi}_\m\gamma_\sigma\gamma_5 D_{\nu}[\omega(e^c_\a,\psi_\a)](\delta_{\epsilon}\psi_{\rho})]=
}\\[4pt]
{\phantom{\delta S_{RS}}=\frac{i}{2\kappa}\!\!\idx\,(\overline{\epsilon}\gamma^\m\psi_a ) e^{\nu}_b R^{ab}_{\m\n}[\omega(e^c_\a,\psi_\a)]-\delta\, S_{Aux}}\\[4pt]
{\delta\, S_{Aux}=-\frac{2}{3}\!\!\idx
	\,\big[S\delta_{\epsilon}S+P\delta_{\epsilon}P+A^{a}\delta_{\epsilon}A_{a}\big].}
\end{array}
\end{equation*}
Notice that in the previous equations we have used that
\begin{equation*}
\gamma_a\psi_\mu e^{a\mu}=0.
\end{equation*}
Otherwise (\ref{vanishSUSG}) would not hold.

Let us move on and compute the commutator of two unimodular supergravity transformations as defined in (\ref{unisugratrans}). Let $\epsilon_1$ and $\epsilon_ 2$ be any two solutions to (\ref{preservingRSGdevelop}), then, the following equations hold
\begin{equation}
\delta_{\epsilon_1}\big(\gamma^\mu\psi_\mu\big)=0,\quad \delta_{\epsilon_2}\big(\gamma^\mu\psi_\mu\big)=0,\quad \delta_{\epsilon_1}\delta_{\epsilon_2}\big(\gamma^\mu\psi_\mu\big)=0.
\label{keyequationtwo}
\end{equation}
Let us point out that $\delta_{\epsilon_1}\delta_{\epsilon_2}\big(\gamma^\mu\psi_\mu\big)=0$ comes from the fact that, in the formal series expansion in $\kappa$ we are using to solve (\ref{preservingRSGdevelop}),
\begin{equation*}
\delta_{\epsilon_2}\big(\gamma^\mu\psi_\mu\big)=F[\epsilon_2(e^a_\mu,\psi_\mu,S,P,A_a),e^a_\mu,\psi_\mu,S,P,A_a]=0
\end{equation*}
holds for any value of $e^a_\mu,\psi, S, P, A_a$ with the appropriate regularity and boundary behaviour --the paragraph right below (\ref{preservingRSGdevelop}) is most relevant in this regard. We have introduced the function $F$ to remark that $\delta_{\epsilon_2}\big(\gamma^\mu\psi_\mu\big)$ depends on the fields both explicitly and implicitly, the latter dependence through $\epsilon_2$.

Using (\ref{keyequationtwo}) and the definitions in (\ref{standarsugratrans}) and (\ref{unisugratrans}), one readily comes to the conclusion that
\begin{equation}
\begin{array}{l}
{\delta_{\epsilon_1}\delta_{\epsilon_2}\big(\text{ufield}\big)=\delta_{\Delta_{12}}
\big(\text{ufield}\big)+[\tilde{\delta}_{\tilde{\epsilon}_1}\tilde{\delta}_{\tilde{\epsilon}_2}\big(\text{field}\big)]_{[\tilde{\epsilon}_1=\epsilon_ 1,\tilde{\epsilon}_2=\epsilon_ 2,\tilde{e}^a_{\mu}=e^a_{\mu},
\tilde{\psi}_{\mu}=\psi_{\mu}]}}\\[4pt]
{\Delta_{12}=\delta_{\epsilon_1}\epsilon_2,\quad\text{ufield}=e^a_\mu,\;\psi_\mu,\; S,\;P,\;A_a;\quad\text{field}=\tilde{e}^a_{\mu},\;\tilde{\psi}_{\mu},\;S,\;P,\;A_a.}
\end{array}
\label{anotherkeyeq}
\end{equation}
Recall that $\epsilon_1$ and $\epsilon_2$ depend on the unimodular fields so that $\Delta_{12}=\delta_{\epsilon_1}\epsilon_2$ is not zero.

By employing the results in (\ref{anotherkeyeq}), one shows that
\begin{equation}
[\delta_{\epsilon_1},\delta_{\epsilon_2}](\text{ufield})=\delta^{(Diff)}_{\xi}(\text{ufield})+\delta^{(Lorentz)}_{\Lambda}(\text{ufield})+\delta_{\Sigma}(\text{ufield}),
\label{closedalgebra}
\end{equation}
where $\delta^{(Diff)}_{\xi}$ is a diffeomorphism with parameters $\xi^a$, $\delta^{(Lorentz)}_{\Lambda}$ denotes a Lorentz transformation with parameters $\Lambda^a_{\phantom{a}b}$ and $\delta_{\Sigma}$ is given by the  supergravity transformations in (\ref{unisugratrans}) with parameter $\Sigma$ instead of $\epsilon$.
The value of each of these parameters is given next:
\begin{equation}
\begin{array}{l}
{\xi^\mu=\frac{i}{2}\overline{\epsilon}_1\gamma^\mu\epsilon_2,}\\[4pt]
{\Lambda^a_{\phantom{a}b}=\xi^\rho\omega_{\rho\phantom{a}b}^{\phantom{\rho}a}+\frac{\kappa}{6}\overline{\epsilon}_2\gamma^a_{\phantom{a}b}(S-i\gamma_5 P)\epsilon_1
-\frac{\kappa}{12}\overline{\epsilon}_2\{\gamma^a_{\phantom{a}b},\gamma^c\}\gamma_5\epsilon_1 A_c,}\\[4pt]
{\Sigma=\delta_{\epsilon_1}\epsilon_2-\delta_{\epsilon_2}\epsilon_1-\kappa \xi^{\rho}\psi_{\rho}.}
\end{array}
\label{parameters}
\end{equation}

Since we want the commutator of two unimodular supergravity transformations in (\ref{unisugratrans}) to yield a unimodular supergravity transformation modulo a transverse diffeomorphism --not a general  diffeomorphism-- and a Lorentz transformation, the equation in (\ref{closedalgebra}) cannot be the end of the story. It remains to be shown that $\partial_{\mu}\xi^\mu=0$ and that $\delta_{\Sigma}(\gamma^\mu \psi_{\mu})=0$. We do so next.

Let us show first that $\partial_{\mu}\xi^\mu=0$, where $\xi^\mu$ is given in (\ref{parameters}). Let $\Gamma^{\rho}_{\mu\nu}(e^a_\sigma,\psi_\sigma)$ be given by
\begin{equation}
\Gamma^{\rho}_{\mu\nu}(e^a_\sigma,\psi_\sigma)=\Gamma^{\rho}_{\mu\nu}(g)-K^{\quad\rho}_{\mu\nu}(e^a_{\sigma},\psi_\sigma),
\label{Chrispsi}
\end{equation}
where $\Gamma^{\rho}_{\mu\nu}(g)$ detones the Christoffel symbols for the unimodular metric $g_{\mu\nu}=e^a_{\mu}e_{a\nu}$ and $K^{\quad\rho}_{\mu\nu}(e^a_{\sigma},\psi_\sigma)$
is defined by
\begin{equation*}
K^{\quad\rho}_{\mu\nu}(e^a_{\sigma},\psi_\sigma)=i\frac{\kappa^2}{4}\big(\overline{\psi}_{\mu}\gamma^\rho \psi_{\nu}-\overline{\psi}_{\nu}\gamma_\mu \psi^{\rho}+\overline{\psi}^{\rho}\gamma_\nu \psi_{\mu}\big).
\end{equation*}
Then, taking advantage of the following results
\begin{equation*}
\begin{array}{l}
{\gamma^{\mu}D_{\mu}[\omega(e^a_{\sigma},\psi_{\sigma})]\epsilon_1=-i\frac{\kappa^2}{2} (\overline{\epsilon_1}\gamma^b\psi_a)\gamma^a\psi_b
-i\frac{2\kappa}{3}(S-i\gamma_5 P)\epsilon_1
-i\frac{\kappa}{6}\gamma_5\gamma^{\nu}\epsilon_1 A_{\nu},}\\[4pt]
{D_{\mu}[\omega(e^a_{\sigma},\psi_{\sigma})]\overline{\epsilon}_2\gamma^\mu=-i\frac{\kappa^2}{2} (\overline{\epsilon}_2\gamma^b\psi_a)\overline{\psi}_b\gamma^a
+i\frac{2\kappa}{3}\overline{\epsilon}_2 (S-i\gamma_5 P) +i\frac{\kappa}{6}\overline{\epsilon}_2\gamma_5\gamma^{\nu}A_{\nu},}\\[4pt]
{\partial_{\mu}\gamma^{\nu}+\frac{1}{4}\omega_{\mu ab}(e^a_{\sigma},\psi_{\sigma})[\gamma^{ab},\gamma^\nu]+\Gamma^{\nu}_{\mu\rho}(e^a_\sigma,\psi_\sigma)\gamma^{\rho}=0,}
\end{array}
\end{equation*}
one shows that
\begin{equation}
 2\partial_{\mu}\xi^\mu=i \overline{\epsilon}_1\gamma^\rho\epsilon_2\,\Gamma^{\mu}_{\mu\rho}(e^a_\sigma,\psi_\sigma).
 \label{tobezero}
\end{equation}
Now, $\Gamma^{\mu}_{\mu\rho}(g)=0$, for $g_{\mu\nu}$ is unimodular, and $K^{\mu}_{\mu\rho}(e^a_\sigma,\psi_\sigma)=0$, since $\psi_{\mu}$ is Majorana and $\gamma^\mu \psi_\mu=0$. Hence, putting together (\ref{Chrispsi}) and (\ref{tobezero}), one reaches the conclusion that $\partial_{\mu}\xi^\mu=0$; as required.

Let us show next that $\delta_{\Sigma}(\gamma^\mu\psi_\mu)=0$, where $\Sigma$ is given in (\ref{parameters}), ie, $\Sigma$ is an admissible parameter for the unimodular supergravity transformation in (\ref{unisugratrans}). Now, since  the equations in (\ref{keyequationtwo}) hold, we have
\begin{equation}
[\delta_{\epsilon_1},\delta_{\epsilon_2}](\gamma^\mu\psi_{\mu})=\gamma^a([\delta_{\epsilon_2},\delta_{\epsilon_1}]e_a^\mu)\psi_{\mu}+\gamma^\mu([\delta_{\epsilon_1},\delta_{\epsilon_2}]\psi_{\mu})=0.
\label{anullresult}
\end{equation}
Using (\ref{closedalgebra}), one can readily deduce that
\begin{equation}
[\delta_{\epsilon_1},\delta_{\epsilon_2}]e_a^\mu=-e^\mu_b e^\nu_a\big(\xi^\rho\partial_\rho e^b_\nu+\partial_\nu\xi^\rho e^b_\rho+\Lambda^b_{\phantom{b}c}e^c_\nu-i\frac{\kappa}{2}
\overline{\Sigma}\gamma^b\psi_\nu\big).
\label{doblevareamu}
\end{equation}
We also have --see (\ref{closedalgebra})-- that
\begin{equation}
\begin{array}{l}
{[\delta_{\epsilon_1},\delta_{\epsilon_2}]\psi_{\mu}=\xi^\rho\partial_\rho\psi_\mu+\partial_\mu\xi^\rho \psi_\rho+\frac{1}{4}\Lambda_{ab}\gamma^{ab}\psi_\mu+
\frac{1}{\kappa}D_{\mu}[\omega^{ab}(e^c_{\mu},\psi_{\mu})]\Sigma}\\[4pt]
{\phantom{[\delta_{\epsilon_1},\delta_{\epsilon_1}]\psi_{\mu}=}
+\frac{i}{6}\gamma_{\mu}(S-i\gamma_5 P)\Sigma +\frac{i}{2}\gamma_5(\delta^{\nu}_{\mu}-\frac{1}{3}\gamma_{\mu}\gamma^{\nu})\Sigma  A_{\nu}}.
\end{array}
\label{doblevarpsimu}
\end{equation}
By substituting first (\ref{doblevareamu}) and (\ref{doblevarpsimu}) in (\ref{anullresult}) and, then, performing a lengthy algebra, one obtains -due to the occurrence of a surprising bunch of cancellations-- that $\Sigma$ satisfies the following equation
\begin{equation*}
\gamma^{\mu}D_{\mu}[\omega(e^a_{\sigma},\psi_{\sigma})]\Sigma=-i\frac{\kappa^2}{2} (\overline{\Sigma}\gamma^b\psi_a)\gamma^a\psi_b
-i\frac{2\kappa}{3}(S-i\gamma_5 P)\Sigma -i\frac{\kappa}{6}\gamma_5\gamma^{\nu}\Sigma A_{\nu}.
\end{equation*}
This is what was required for $\Sigma$ to be an admissible unimodular supergravity transformation parameter.

Summarizing, we have shown that the unimodular supergravity transformations in (\ref{unisugratrans}) form a closed algebra modulo transverse diffeomorphims and Lorentz transformations.

\newpage

\section{ Unimodular supergravity and its classical solutions}
A classical solution of a supergravity theory is a bosonic field configuration which satisfies
the equations of motion of the theory once the auxiliary fields have been removed and the
fermionic fields have been set to zero. Not every classical solution of a supergravity theory is
invariant under the supergravity transformations defining the latter. Those classical solutions
that preserve some of the aforementioned supergravity transformations are called supersymetric
or BPS solutions.
In the case of the unimodular supergravity theory whose action is in (3.1), setting $\psi_\mu =0$
and $S = 0 , P = 0$ and $A_\mu = 0$ leads to the conclusion that the classical solutions of the
theory at hand are those unimodular metrics which are solutions to the unimodular gravity
equations of motion, which read:
\begin{equation}
R_{\mu\nu}-\frac{1}{4} R g_{\mu\nu}=0.
\label{unieq}
\end{equation}
$R_{\mu\nu}$ and $R$ denote the Ricci tensor and the scalar curvature for the unimodular metric
$g_{\mu\nu}= e^a_\mu e_{a\nu}$ , respectively.
It is plain that any classical solution of the standard $N = 1, d = 4$ Poincar\'e supergravity
equation of motion is a solution to the equations of motion of the unimodular supergravity theory put forward in this paper, but not the other way around.
Indeed, the equations in (\ref{unieq}) admit de Sitter and anti-de Sitter spacetimes as solutions; these
spacetimes are not classical solutions of standard $N = 1, d = 4$ Poincar\'e supergravity. So,
standard $N = 1, d = 4$ Poincar\'e supergravity is not equivalent to the unimodular gravity
theory whose action is given (\ref{SUSG}) in the sense that their spaces of classical solutions are not the
same. However, the solutions to (\ref{unieq}) with a non vanishing Cosmological Constant are never invariant under the unimodular transformations  (\ref{unisugratrans}). Indeed, the supergravity invariance condition
\begin{equation*}
0=\delta_{\epsilon}\psi_{\mu}=\frac{1}{\kappa}D_\mu\epsilon,\quad D_{\mu}=\partial_{\mu}+\frac{1}{4}\omega^{ab}_\rho(e^c_\mu)\gamma_{ab},
\end{equation*}
where $\omega^{ab}_\rho(e^c_\mu)$ is the Levi-Civita spin connection, implies that $\epsilon$ has to be a killing spinor. But it is known \cite{Bohle:2003abk} that if such a spinor would exist then $R=0$, which would contradict the hypothesis of a non vanishing Cosmological Constant. Notice that when $\psi_\mu=0$, $S=0$, $P=0$ and $A_a=0$, any killing spinor satisfies (\ref{preservingRSGdevelop}) --which boils down to $\gamma^\mu D_\mu[\omega^{ab}_\rho(e^c_\mu)]\epsilon=0$-\ and, therefore, it is an admissible unimodular supergravity transformation parameter. We stress that, unlike in the standard $N=1,d=4$ Poincar\'e supergravity case, both de Sitter and anti-de Sitter are vacua of the unimodular $N=1, d=4$ theory; they break --spontaneously-- unimodular supergravity invariance, though.

It is well known \cite{Ortin:2015hya} that the maximally supersymmetric vacuum of standard $N=1, d=4$ is Minkowski spacetime. This spacetime is also the unique maximally supersymmetric vacumm of unimodular $N=1, d=4$ Poincar\'e supergravity, for any killing spinor satifies (\ref{preservingRSGdevelop}) when $\psi_\mu=0$, and, of course, $S=0$, $P=0$, $A_a=0$. Hence, it is legitimate to view unimodular $N=1, d=4$ supergravity as a  theory of gravitons and gravitinos propagating in Minkowski spacetime, in the perturbative $\kappa$-expansion.

Finally, it is plain that if we look for classical solutions of our $N=1, d=4$  unimodular supergravity which partially break unimodular supergravity, we shall only find the standard  gravitational pp-waves: one has to solve the same killing spinor equation \cite{Ortin:2015hya}  as in the ordinary $N=1, d=4$  Poincar\'e supergravity.

\section{Conclusions}

The main conclusion of this paper is that a unimodular $N = 1, d = 4$ Poincar\'e supergravity
can be formulated off-shell. This unimodular gravity theory is the counterpart of the standard
$N = 1, d = 4$ Poincar\'e supergravity. Analogously to the case of unimodular gravity, the infinitesimal parameters defining the unimodular supergravity
transformations are constrained by a differential equation which make those parameters field
dependent. Indeed, along with the gravitational unimodular contraint $ \det e^a_\mu=1$, one has
to impose the constraint $\gamma^\mu\psi_\mu= 0$, if e = 1 is to be preserved by the supergravity transformations.
That the constraint $\gamma^\mu\psi_\mu= 0$ be preserved under supegravity transformations leads
to the differential equation we have just mentioned. We have shown that despite that field dependence
the commutator of two unimodular supergravity transformations closes on transverse
diffeomorphisms, Lorentz transformations and unimodular supergravity transformations.
We have shown that the unimodular $N = 1, d = 4$ Poincar\'e supergravity theory presented
here has de Sitter and anti-de Sitter spacetimes as classical solutions that break -spontaneously-- supergravity.
This phenomenon makes unimodular $N = 1, d = 4$ supergravity different from its standard
counterpart $N = 1, d = 4$ Poincar\'e supergravity, for all classical solutions to
the latter are Ricci flat. And yet, the only maximally supersymmetric vacuum solution of our
unimodular supergravity theory is Minkowski spacetime. Around this vacuum our theory is
a theory of interacting gravitons and gravitinos as we have shown by studying the linearized
unimodular supergravity theory which has a (rigid) N = 1 supersymmetry.
Finally, it seems clear that the unimodular counterpart of anti-de Sitter supergravity can
be formulated by generalizing the framework presented here along the lines of Reference [11].
Thus we will have a unimodular theory whose maximally superymmetric vacuum is anti-de
Sitter spacetime.

\section{ Acknowledgements}

We are very much indebted to E. \'Alvarez for countless discussions on unimodular supergravity
and for guidance. We thank T. Ort\'{\i}n for provinding us with information relevant to this
paper.  This work has received funding from the Spanish Research Agency (Agencia Estatal de Investigacion) through the grant IFT Centro de Excelencia Severo Ochoa SEV-2016-0597, and the European Union's Horizon 2020 research and innovation programme under the Marie Sklodowska-Curie grants agreement No 674896 and No 690575. We also have been partially supported by FPA2016-78645-P(Spain). RSG is supported by the Spanish FPU Grant No FPU16/01595. CPM has been partially supported by the Spanish Ministerio de Ciencia, Innovaci\'on y
Universidades through grant PGC2018-095382-B-I00.

%%%%%%%%%%%%%%%%%%%%%%%%%%%%%%%%%%%%%%%%%%%%%%%%%%%%%%%%
%%%%%%%%%%%%%%%%%%%%%%%%%%%%%%%%%%%%%%%%%%%%%%%%%%%%%%%%%
%%%%%%%%%%%%%%%%%%%%%%%%%%%%%%%%%%%%%%%%%%%%%%%%%%

\end{document}